\documentclass[preprint2]{aastex631}
\begin{document}
\title{Strengthening the Link Between Fullerenes and a Subset of Diffuse Interstellar Bands}

\author[0000-0001-8803-3840]{Daniel Majaess}
\affiliation{Department of Chemistry and Physics, Mount Saint Vincent University, Halifax, Nova Scotia, B3M2J6 Canada.}
\email{Daniel.Majaess@msvu.ca}

\author[0000-0003-3469-8980]{Tina A. Harriott}
\affiliation{Department of Mathematics and Statistics, Mount Saint Vincent University, Halifax, Nova Scotia, B3M2J6 Canada.}
\affiliation{Department of Chemistry and Physics, Mount Saint Vincent University, Halifax, Nova Scotia, B3M2J6 Canada.}

\author{Halis Seuret}
\affiliation{Centro de Investigaciones Químicas, IICBA, Universidad Autónoma del Estado de Morelos, Cuernavaca, 62209, Morelos, Mexico.}
\affiliation{Department of Chemistry and Physics, Mount Saint Vincent University, Halifax, Nova Scotia, B3M2J6 Canada.}

\author[0000-0002-8746-9076]{Cercis Morera-Boado}
\affiliation{IXM-Cátedra Conahcyt-Centro de Investigaciones Químicas, IICBA, Universidad Autónoma del Estado de Morelos, Cuernavaca, 62209, Morelos, Mexico.}

\author[0000-0001-6662-3428]{Lou Massa}
\affiliation{Hunter College \& the PhD Program of the Graduate Center, City University of New York, New York, USA.}

\author[0000-0001-8397-5353]{Ch\'erif F. Matta}
\affiliation{Department of Chemistry and Physics, Mount Saint Vincent University, Halifax, Nova Scotia, B3M2J6 Canada.}
\affiliation{Department of Chemistry, Saint Mary's University, Halifax, Nova Scotia, B3H3C3 Canada.}
\affiliation{D\'epartement de Chimie, Universit\'e Laval, Qu\'ebec, G1V0A6 Canada.}
\affiliation{Department of Chemistry, Dalhousie University, Halifax, Nova Scotia, B3H4J3 Canada.}

\begin{abstract}
A debate persists regarding the correlation between the DIBs 9577 and 9632 {\AA}, and whether they share a common molecular carrier (i.e., C$_{60}^{+}$).  A robust high correlation determination emerges after bridging the baseline across an order of magnitude ($\simeq 50 - 700$ m{\AA}, $r=0.93\pm0.02$), and nearly doubling the important higher equivalent width domain by adding new Mg II-corrected sightlines.  Moreover, additional evidence is presented of possible DIB linkages to fullerenes, whereby attention is drawn to DIBs at 7470.38, 7558.44, and 7581.47 {\AA}, which match the Campbell experimental results for C$_{70}^{+}$ within 1 {\AA}, and the same is true of 6926.48 and 7030.26 {\AA} for C$_{70}^{2+}$. Yet their current correlation uncertainties are unsatisfactory and exacerbated by expectedly low EWs (e.g., $\overline{EW}=4$ m{\AA} for 6926.48 {\AA}), and thus further observations are required to assess whether they represent a \textit{bona fide} connection or numerical coincidence.
\end{abstract}

\section{Introduction}

\begin{figure*}
\begin{center}
\includegraphics[width=0.8\linewidth]{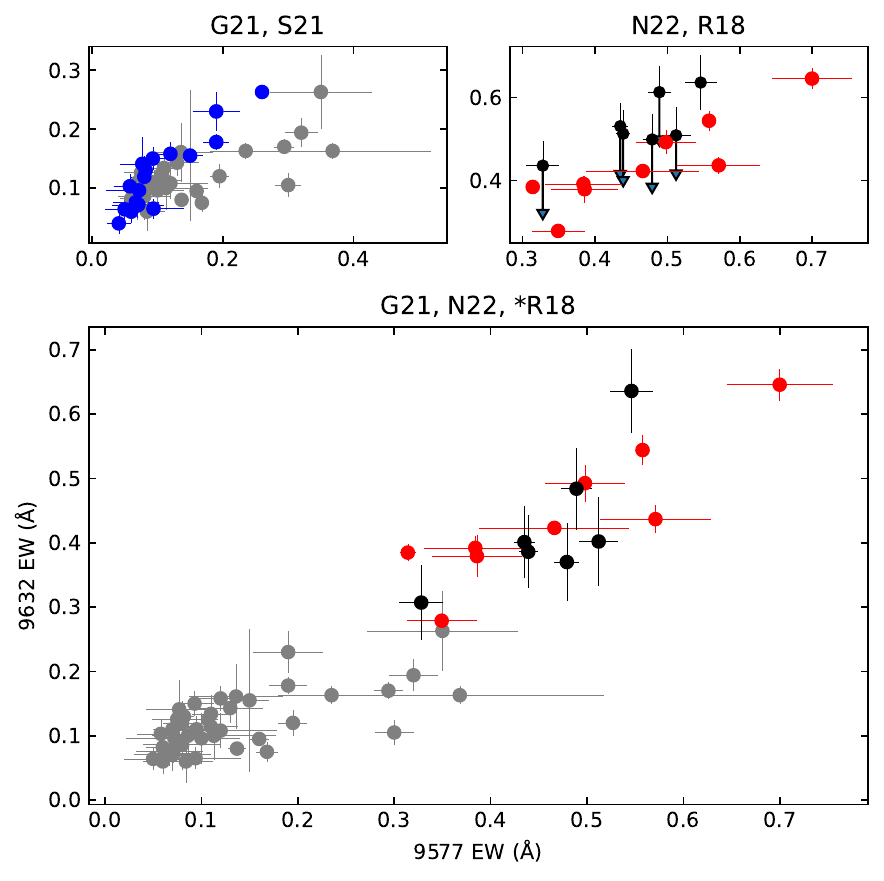}
\caption{DIBs 9577 and 9632 {\AA} are characterized by a high correlation across a sizable $\simeq 650$ m{\AA} EW baseline ($r=0.93\pm0.02$, G21, N22, *R18).  References correspond to \citet[][gray]{gal21}, \citet[][blue]{sch21}, and \citet[][red]{nie22}. Advantageous higher EW sightlines were added from \citet[][black]{rt18} since they mostly do not overlap with \citet{nie22}, and were adjusted here for stellar Mg II contamination (bottom panel, *R18, see text). Arrows in the top right panel convey the Mg II corrections.  Table~\ref{table:summary} highlights the dependence of $r$ and the slope on permutations of the underlying data.}
\label{fig1}
\end{center}
\end{figure*}

\citet{gal17a,gal21}, \citet{sch21}, and \citet{nie22} debated the correlation between the 9577 and 9632 {\AA} DIBs, which experiments highlight as prominent absorption features associated with C$_{60}^{+}$ \citep[e.g.,][]{fu93,ca15}.  \citet{gal21} argued those DIBs exhibit too low a correlation to be associated, or an unknown causes the correlation to vary (e.g., blended lines).  Conversely, \citet{sch21} and \citet{nie22} favored a high correlation and unambiguous link to C$_{60}^{+}$.  DIBs tied to 9365.2 and 9427.8 {\AA} may also be associated with C$_{60}^{+}$ \citep{ca16,wa16,la18,co19}, although here too the conclusion is contested \citep{gal21}.

The topic is pertinent since C$_{60}^{+}$ is hitherto the sole carrier of DIBs where a consensus is coalescing \citep[e.g.,][]{co19,nie22}.  Several hundred other DIBs remain unassociated with specific molecules.  Majaess et al.~(\textit{submitted}) advocated that the signatures of C$-$H, C$=$O, C$\equiv$C, C$\equiv$N, S$-$H, and aromatics ($=$C$-$H stretch, out of plane C$-$H bending, in-ring C$\mathbf{^{\underline{...}}}$C, overtones) are discernible in energy differences between correlated DIBs in the APO catalog \citep{Fan2019}.  Indeed, PAHs are hypothesized to be leading candidates for DIB sources \citep[e.g.,][]{bon20,web22}, and efforts are likewise ongoing to explore the viability of fullerenes beyond C$_{60}^{+}$ (\S \ref{sec:c70}), heterofullerenes, and their (endo/exo)hedral inclusions \citep[e.g.,][]{ca16b,om16}.  A subset of those molecules may likewise explain the unidentified infrared emission lines \citep{sa20,sa22}, in addition to MAONs \citep[mixed aromatic/aliphatic organic nanoparticles,][]{kw22,ks23}. 

In this study, concerns regarding the correlation between the 9577 and 9632 {\AA} DIBs are assuaged.  Furthermore, additional DIBs are inspected whose source may also be connected to fullerenes (i.e., C$_{70}^{+}$ and C$_{70}^{2+}$), given tightly constrained matching wavelengths \textit{vis \`a vis} laboratory results \citep{ca16b,ca17}. 

\section{Analysis}
\subsection{C$_{60}^{+}$: 9577 and 9632 {\AA}}
  \citet{gal21} concluded that the correlation between 9577 and 9632 {\AA} is $r=0.37$, whereas \citet{sch21} favored estimates spanning $0.82-0.93$.  \citet{nie22} inspected separate stars featuring higher EWs and advocated that $r=0.89-0.96$.
  
  It is argued here that the ambiguity partly arises from not bridging the analysis over a sizable EW baseline, whereby \citet{gal21} and \citet{sch21} sample relatively low EWs, \citet{nie22} data reside within a higher EW domain, and linking the datasets reveals a robust high correlation between 9577 and 9632 {\AA} (Fig.~\ref{fig1}, Table~\ref{table:summary}). \citet[][their Table 1]{sch21} relied principally on \citet{gal21} data (Fig.~\ref{fig1}, top left panel), and consequently the former is omitted from the final correlation determination (bottom panel of Fig.~\ref{fig1}, $r=0.93\pm0.02$).

\begin{deluxetable}{ccc}
\tablecaption{Correlation and slope for 9577-9632 {\AA}.\label{table:summary}}
\tablehead{\colhead{Datasets} & \colhead{$r\pm \Delta r$} & \colhead{$m\pm \Delta m$}}
\startdata
G21, N22 & $0.92\pm0.02$ & $0.80\pm0.04$ \\
S21, N22 & $0.98\pm0.01$ & $0.82\pm0.05$ \\
G21, *R18 & $0.89\pm0.03$ & $0.77\pm0.06$ \\ 
S21, *R18 & $0.96\pm0.02$ & $0.86\pm0.05$ \\
G21, N22, *R18 & $0.93\pm0.02$ & $0.83\pm0.04$ \\
S21, N22, *R18 & $0.97\pm0.01$ & $0.86\pm0.04$ \\
G21, S21, N22, *R18 & $0.94\pm0.01$ & $0.83\pm0.04$ \\
\enddata
\tablenotetext{ }{Notes:~references are \citet[][G21]{gal21}, \citet[][S21]{sch21}, and \citet[][N22]{nie22}. Sightlines linked to comparatively cooler stars were added since they are mostly not featured in the other datasets, and once corrected for stellar Mg II contamination (\citealt{rt18}, *R18) (see text).}
\end{deluxetable}

  Permutations of the datasets in concert with their slope and correlation determinations are conveyed in Table~\ref{table:summary}, which stemmed from an unweighted evaluation since the uncertainties are inhomogeneous. A firm high correlation remains in the absence of the \citet{sch21} data, or the adjusted \citet{rt18} findings (Table~\ref{table:rt18}). Regarding the latter, pertinent high EW datapoints were added from \citet{rt18} once their uncorrected 9632 {\AA} EWs were adjusted for stellar Mg II contamination. Those stars largely differ from the subsample presented by \citet{nie22} (B215 overlaps), who focused on hotter stars to mitigate the impact of Mg II.  Tables~1 and 2 in \citet{gal17a} include corrections for stars spanning diverse parameters (e.g., $T_{\rm eff}$, metallicity, $v_{T}$, $\log{g}$), and a sigmoid was applied to approximate an upper bound Mg II$-T_{\rm eff}$ relation over the baseline examined: ${\Delta EW \approx (130\pm50 \; {\rm m\text{\AA}})(1-(1+e^{(20200-T_{\rm eff})/1751})^{-1})}$.  Corrections emerging from that approximation were subsequently applied to the seven stars using spectral types from \citet{rt17,rt18},\footnote{A B8 temperture class was assumed when \citet{rt18} relayed a `late-B' classification.} and their corresponding temperatures stemmed from unpublished data by D.~G.~Turner \citep[e.g., used in][and references therein]{tu94}. Uncertainties for the corrected 9632 {\AA} \citet{rt18} EWs were all expanded in quadrature by a bulk 50 m{\AA} from their original values. Table~\ref{table:rt18} underscores that the corrections shift the original \citet{rt18} data downward in Fig.~\ref{fig1} (top right panel) and upon the \citet{nie22} observations. Otherwise, the initial \citet{rt18} EWs are too high, and discernibly offset from the \citet{nie22} data.  As noted \citet{nie22} do not require corrections owing to their hotter temperature sample. That in sum reaffirms that the approximated Mg II corrections are satisfactory for the present purpose, despite the uncertainties (e.g., deviations from the mean trend, SpT, $T_{\rm eff}$, possible Doppler offsets between Mg II and the DIB, and for the latter see Fig.~4 in \citealt{gal17a}). \citet{la18} and \citet{gal21} debated the Mg II corrections employed by \citet{gal17a}, and for additional approaches to the problem consider \citet{jen97} and \citet{wa16}. A comprehensive spectral analysis comparing differing atmospheric models is desirable but beyond the scope of this current effort, especially given the scaffolding underpinning Table~\ref{table:summary} and the consistently high correlation conclusion. 

\begin{deluxetable}{lccc}
\tablecaption{\citet{rt18} original and corrected 9632 {\AA} EWs.\label{table:rt18}}
\tablehead{\colhead{ID} & \colhead{SpT} & \colhead{EW$_0$ (m{\AA})} & \colhead{EW$_{\rm c}$ (m{\AA})}}
\startdata
B215 & B0-B1V & 636$\pm$42 &	636$\pm$65 \\
B243 & B8V & 499$\pm$34 &	370$\pm$60 \\
B253 & B3-B5III & 509$\pm$47 &	402$\pm$69 \\
B268 & B9-A0 & 531$\pm$25 &	401$\pm$56 \\
B275 & B7III & 513$\pm$27 &	386$\pm$57 \\
B331 & late-B & 436$\pm$31 &	307$\pm$59 \\
B337 & late-B & 613$\pm$39 & 484$\pm$63 \\
\enddata
\tablenotetext{ }{Notes:~IDs and spectral types stem from \citet{rt18}. Their 9632 {\AA} EWs were corrected (EW$_{\rm c}$) for stellar Mg II contamination (see text).  A high correlation between 9632 and 9577 {\AA} persists in the absence of the aforementioned data (Table~\ref{table:summary}).}
\end{deluxetable}

\begin{deluxetable}{cccc}
\tablecaption{\textit{Candidate} fullerene DIBs.\label{table:dibfamily}}
\tablehead{\colhead{DIB ({\AA})} & \colhead{$\lambda_e$ ({\AA})} & \colhead{Ion} & \colhead{$\lambda_e$ (ref.)}}
\startdata
7470.38 & 7470.2 & C$_{70}^{+}$ & \citealt{ca16b} \\
7558.44 & 7558.4 & C$_{70}^{+}$ & \citealt{ca16b} \\
7581.47 & 7582.3 & C$_{70}^{+}$ & \citealt{ca16b} \\
\hline 
6926.48 & 6927 & C$_{70}^{2+}$ & \citealt{ca17} \\
7030.26 & 7030 & C$_{70}^{2+}$ & \citealt{ca17} \\
\enddata
\tablenotetext{ }{Notes:~DIB wavelengths stem from the APO catalog \citep{Fan2019}, save 7558.44 {\AA} \citep[][see text]{jd94,ho09}.  $\lambda_e$ are experimental wavelengths \citep{ca16b,ca17}.}
\end{deluxetable}

\begin{deluxetable}{ccccccc}
\tablecaption{DIB correlations, slopes, and lab attenuation ratios.\label{table:corr}}
\tablehead{\colhead{DIB} & \colhead{$\overline{EW}$} & \colhead{DIB} & \colhead{$n$} & \colhead{$r$} & \colhead{$m$} & \colhead{$R_{a}$} \\ 
\colhead{({\AA})} & \colhead{(m{\AA})} & \colhead{({\AA})} & \colhead{} & \colhead{$\pm \Delta r$} & \colhead{$\pm \Delta m$} & \colhead{$\pm \Delta R_a$}}
\startdata
C$_{70}^{+}$ & & & & & & \\
7470.38 & 6 & 7557.88 & 12 & $0.84$ & $2.6$ & $1.0$ \\
& & & & $\pm0.10$ & $\pm0.5$ & $\pm0.3$ \\
7470.38 & 5 & 7581.47 & 17 & $0.80$ & $4.2$ & $1.3$ \\
& & & & $\pm0.10$ & $\pm0.8$ & $\pm0.4$ \\
7557.88 & 12 & 7581.47 & 14 & $0.76$ & $1.6$ & $1.3$  \\
& & & & $\pm0.13$ & $\pm0.4$ & $\pm0.4$ \\
\hline 
C$_{70}^{2+}$ & & & & & & \\
6926.48 & 4 & 7030.26 & 9 & $0.84$ & $0.8$ & $2.0$ \\ 
& & & & $\pm0.12$ & $\pm0.2$ & $\pm0.6$ \\
\enddata
\tablenotetext{ }{Notes:~DIB wavelengths and EWs from the APO catalog \citep{Fan2019}. Low statistics and EWs, and a limited baseline, suggest the  uncertainty is larger than cited.}  
\end{deluxetable}

\begin{figure}
\begin{center}
\includegraphics[width=0.99\linewidth]{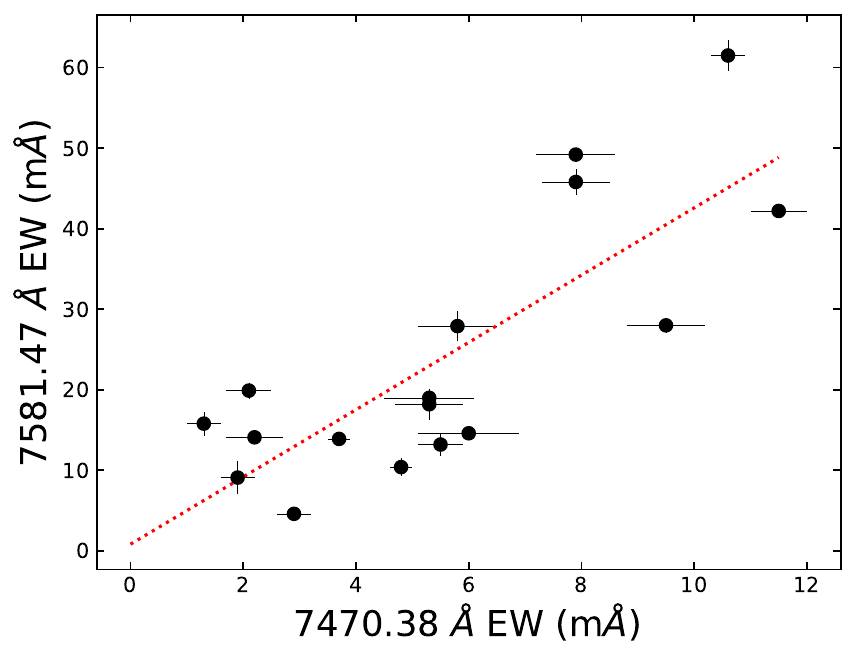}
\caption{Correlation analysis between two DIBs ($7470.38-7581.47$ {\AA}) which exhibit wavelengths comparable to experimental C$_{70}^{+}$ lines (Table~\ref{table:dibfamily}). The expected low EWs \citep[$\lambda_e=7470.38$ {\AA},][]{ca16b} help expand the correlation uncertainty ($r=0.80\pm0.10$).  The slope ($m=4.2\pm0.8$) is larger than an indirect comparison to the relative lab-measured attenuation ($R_a$, Table~\ref{table:corr}).}
\label{fig:corr}
\end{center}
\end{figure}

\subsection{The {\rm C}$_{70}^{+}$ and {\rm C}$_{70}^{2+}$ DIB families}
\label{sec:c70}
There may be DIBs tied to other fullerenes (e.g., C$_{70}^{+}$, C$_{70}^{2+}$, Table~\ref{table:dibfamily}).  Those lines could help demarcate mid-infrared vibrations, whereby energy differences among DIBs within the family (e.g., highly correlated EWs) could represent separate vibrational modes.  Experimental wavelengths for C$_{70}^{+}$ and C$_{70}^{2+}$ were adopted from \citet{ca16b,ca17}, and compared to published DIB wavelengths \citep{jd94,ho09,bon12,Fan2019}. For C$_{70}^{+}$ the following DIBs align with experimental results to within 1 {\AA}: 7470.38, 7558.44, and 7581.47 {\AA}. \citet{ca16b} relay that the most prominent laboratory C$_{70}^{+}$ line ($7632.6$ {\AA}) shall be severely contaminated by telluric absorption, and hence the motivation for satellite measurements \citep[e.g.,][]{co19}. Furthermore, APO catalog DIBs linked to 6926.48 and 7030.26 {\AA} may be indicative of C$_{70}^{2+}$, again owing to their wavelength proximity relative to experimental determinations \citep{ca17}.  Regarding the \citet{ca16b} laboratory C$_{70}^{+}$ wavelength at $\lambda_e=7558.4$ {\AA}, \citet{jd94} and \citet{bon12} feature DIBs at 7558.5 and 7559.19 {\AA}, respectively. That is supported by \citet{ho09}, who characterized DIBs at 7558.44 and 7559.48 {\AA}, with the former being numerically coincident with the experimental result of 7558.4 {\AA} (Table~\ref{table:dibfamily}). The \citet{Fan2019} compilation possesses lines bracketing the laboratory measurement at 7557.88 and 7559.43 {\AA}, which in tandem with the aforementioned \citet{ho09} DIBs, could represent the same vibrational line with differing rotation.\footnote{That may likewise be true of the \citet{Fan2019} 5779.59 and 5780.64 {\AA} DIBs \citep{fra21}.} 

An effort was undertaken to (in)validate the wavelength analysis by one tied to correlations (Table~\ref{table:corr}).  EWs and wavelengths were adopted from \citet{Fan2019} given their sample size (i.e., number of sightlines $n \ge 9$).  Unweighted correlations that emerged include $r=0.84\pm0.10$ for the $7470.38-7557.88$ {\AA} DIB pair (Table~\ref{table:corr}), and $r=0.80\pm0.10$ for $7470.38-7581.47$ {\AA} (Fig.~\ref{fig:corr}).  The correlation evaluations are merely suggestive owing partly to the impact on uncertainties from expectedly\footnote{\citet{ca16b,ca17}.} low EWs (e.g., 7470.38 {\AA} has a median $\overline{EW}=5$ m{\AA}). That uncertainty likewise complicates a comparison between low interstellar EWs and relative laboratory attenuation ($R_a$, Table~\ref{table:corr}), which measure separate marginal (in this instance) quantities in differing environments.  For example, experiments implied that $\lambda_e=7470.2$ {\AA} exhibits a lower intensity than 7582.3 {\AA} \citep{ca16b}, and that is apparent in Fig.~\ref{fig:corr}.  The slope characterizing those DIB EWs ($m=4.2\pm0.8$) is larger than an indirect comparison to the relative lab-measured attenuation ($R_a\sim 1.3$).  As noted above, it is also unclear whether the highly correlated APO catalog lines of 7557.88 and 7559.43 {\AA} ($r=0.92\pm0.05$, $n=16$) constitute a single vibrational line, and the EWs must be combined prior to a comparison with experimental results. Additional observations across a sizable EW baseline may clarify whether the DIBs belong to fullerene carriers, or represent a numerical coincidence.  Yet reliably measuring such low EWs poses a challenge, especially given the experimental FWHM, as noted previously \citep{ca16b,ca17}. 

\section{Conclusions}
DIBs linked to the buckminsterfullerene cation at 9577 and 9632 {\AA} are highly correlated (Fig.~\ref{fig1}, $r=0.93\pm0.02$).  That assessment stemmed from an analysis spanning a lucrative EW baseline, and bridging low to high EWs provides a confident conclusion.  The EWs were tied to existing \citep{gal21,nie22} and newly corrected estimates \citep{rt18}.  Regarding the latter, 9632 {\AA} datapoints linked to relatively cooler stars within \citet{rt18} were corrected here for Mg II contamination (top right panel of Fig.~\ref{fig1}). Their addition reaffirms the overarching conclusion of a high Pearson correlation (Table~\ref{table:summary}).  

The analysis was expanded to identify DIBs associated with fullerenes beyond C$_{60}^{+}$ (i.e., C$_{70}^{+}$ and C$_{70}^{2+}$). Several DIBs overlap with the \citet{ca16b,ca17} laboratory wavelengths to within an Angstrom (Table~\ref{table:dibfamily}, e.g., DIB 7558.44 {\AA} and C$_{70}^{+}$ $\lambda_e=7558.4$ {\AA}). A complimentary correlation analysis was limited by low EWs and statistics (Fig.~\ref{fig:corr}, Table~\ref{table:corr}, e.g., $\overline{EW}=4$ m{\AA} for 6926.48 {\AA}).  Additional observations over an extensive EW baseline are required to assess whether those DIBs are linked to fullerene cations.  The current evidence, though inconclusive, provides sufficient impetus to pursue such observations.  Future work may include relinquishing a strict $<1$ {\AA} criterion, and exploring matches relative to fractionary offsets within a FWHM (e.g., $|{\rm DIB} - \lambda_e |<\frac{1}{x} {\rm FWHM}$).

\begin{acknowledgments}
This research relied on initiatives such as \citeauthor{ca16b}, APO Catalog of DIBs, CDS, NASA ADS, arXiv.
\end{acknowledgments}

\bibliography{article}{}
\bibliographystyle{aasjournal}
\end{document}